\documentclass[11pt]{article}
\usepackage{jheppub}

\usepackage{color}
\usepackage{amsmath}

\usepackage{verbatim}   
\usepackage{subfigure}  
\usepackage{amsfonts}
\usepackage{amssymb}
\usepackage{mathrsfs}
\usepackage{graphicx}
\usepackage{slashed}
\usepackage{amsthm}

\usepackage{graphicx}
\usepackage{dcolumn}
\usepackage{bm}

\usepackage{color}
\usepackage{amsmath}

\usepackage{float}

\usepackage{verbatim}   
\usepackage{subfigure}  
\usepackage{amsfonts}
\usepackage{amssymb}
\usepackage{mathrsfs}
\usepackage{graphicx}
\usepackage{slashed}
\usepackage{amsthm}
\newcommand{\C}{\mathbb{C}}

\def\beq{\begin{equation}\begin{aligned}}
\def\eeq{\end{aligned}\end{equation}}

\def\P{\mathbb{P}}
\def\Z{\mathbb{Z}}
\def\R{\mathbb{R}}
\def\C{\mathbb{C}}

\def\bar#1{\overline{#1}}

\def\inv{^{\raise.15ex\hbox{${\scriptscriptstyle -}$}\kern-.05em 1}}
\def\lbar{{\lower.35ex\hbox{$\mathchar'26$}\mkern-10mu\lambda}} 

\let\<=\langle
\let\>=\rangle

\let\+=\uparrow

\definecolor{mygreen}{RGB}{29,145,70}
\definecolor{mypurple}{RGB}{164,64,214}
\definecolor{myorange}{RGB}{199,146,32}

\newtheorem*{Theorem*}{Theorem}

\definecolor{orange}{rgb}{1,0.5,0}

\begin{document}

\title{Counting Associatives in Compact $G_2$ Orbifolds}

\author[a,b]{Bobby Samir Acharya}
\author[c]{Andreas P. Braun}
\author[a,b]{Eirik Eik Svanes}
\author[d,e,b]{Roberto Valandro}

\affiliation[a]{Department of Physics, Kings College London, London, WC2R 2LS, UK}
\affiliation[b]{Abdus Salam International Centre for Theoretical Physics, Strada Costiera 11, 34151, Trieste, Italy}
\affiliation[c]{Mathematical Institute, University of Oxford, Woodstock Road, Oxford, OX2 6GG, UK}
\affiliation[d]{Dipartimento di Fisica, Universita di Trieste, Strada Costiera 11, 34151 Trieste, Italy}
\affiliation[e]{INFN, Sezione di Trieste, Via Valerio 2, 34127 Trieste, Italy}


\begin{flushright}
KCL-PH-TH/2018-74
\end{flushright}


\abstract{We describe a class of compact $G_2$ orbifolds constructed from non-symplectic involutions of K3 surfaces. Within this class, we identify a model for which there are infinitely many associative submanifolds contributing to the effective superpotential of $M$-theory compactifications. Under a chain of dualities, these can be mapped to $F$-theory on a Calabi-Yau fourfold, and we find that they are dual to an example studied by Donagi, Grassi and Witten. Finally, we give two different descriptions of our main example and the associative submanifolds as a twisted connected sum.}

\maketitle

\section{Introduction}

Supersymmetric theories in four dimensions with minimal supersymmetry encompass an extremely rich class of physical theories in which a huge amount of progress in understanding has been made over the past few decades, and include popular models of particle physics, dark matter and cosmology beyond the Standard Model. Such theories, at low energies, are described by an effective $\mathcal{N}=1$ supergravity model. From a more fundamental, microscopic point of view they can arise by considering $F$-theory on compact elliptic Calabi-Yau fourfolds, heterotic string theories on Calabi-Yau threefolds or, as is the focus of this paper, $M$-theory on compact $G_2$-holonomy spaces. The effective supergravity model is usually specified by a gauge coupling function, $f$, a K\"ahler potential, $K$ and a superpotential, $W$, all of which are functions of the moduli and are determined by the geometry of the Calabi-Yau or $G_2$-holonomy space.

One of the interesting and difficult problems in string/$M$/$F$-theory compactifications with 4D $\mathcal{N}=1$ supersymmetry is the determination of the effective superpotential. While it is relatively straightforward to find the classical superpotential in supergravity, the full theory receives non-perturbative corrections, which can typically be understood as arising from instantonic branes. 
These corrections are often tied to the underlying geometry and topology: for $M$-theory on $G_2$ manifolds, they arise from M2-branes wrapped on associative submanifolds represented by homology spheres \cite{Harvey:1999as}, and for $F$-theory on Calabi-Yau fourfolds they arise from M5-branes wrapped on rigid divisors \cite{Witten:1996bn}. One would therefore like to understand which homology classes can be represented by calibrated cycles. Conversely, knowledge about the existence of non-perturbative physics can provide information about calibrated cycles in the compact internal geometry.

In this paper we will describe some models in which there are an infinite number of contributions to $W$. In particular, we will construct a class of compact $G_2$-holonomy orbifolds containing an infinite number of non-homologous associative sub-orbifolds all of which are diffeomorphic to 3-spheres.

One related example in $F$-theory is due to Donagi, Grassi and Witten~\cite{Donagi:1996yf}. They showed that a particular compact Calabi-Yau four-fold $Y$ contains an infinite number of calibrated six-cycles, corresponding to divisors of $Y$, which give rise to contributions to the effective superpotential. In particular, it was shown in \cite{Donagi:1996yf} that there is a calibrated cycle for every element of the root lattice of $E_8$ and that the naive superpotential is given by an $E_8$ theta function. Using various string dualities, it was then realised that the non-perturbative superpotential of Donagi, Grassi and Witten should have avatars in the other string theories. For instance, under $F$-theory/heterotic duality \cite{Friedman:1997yq},
Curio and L\"ust found the heterotic representative in terms of world-sheet instanton corrections \cite{Curio:1997rn}. Similarly, many heterotic models are expected to have dual descriptions represented by $M$-theory on $K3$-fibred $G_2$-holonomy spaces \cite{Acharya:1996ci,Acharya:2001gy,gukov2003duality}. Indeed, the heterotic/$M$-theory duality map discovered in \cite{Braun:2017uku} was recently applied to these heterotic models \cite{Braun:2018fdp}, and it was shown how the divisors of \cite{Donagi:1996yf} get mapped to associative submanifolds of the $G_2$-holonomy space. These are the associatives that we construct explicitly here.

Instead of following the string duality chain via heterotic to $M$-theory, one can also follow a different path. By taking the type IIB weak coupling limit, the above $F$-theory divisors reduce to calibrated surfaces in the type IIB Calabi-Yau three-fold, which turns out in this case to be the double elliptic Calabi-Yau threefold, $Z$, with $h^{1,1}=h^{2,1}=19$ discovered by Schoen \cite{gukov2003duality}. In fact, all of these surfaces are rational elliptic surfaces and are all divisors in $Z$. Under mirror symmetry, type IIB theory on $Z$ gets mapped to type IIA theory on $\hat{Z}$, the mirror of $Z$, and the divisors of $Z$ are mapped to special Lagrangian 3-cycles in $\hat{Z}$. In this case, $Z$ is self-mirror in the sense that its quantum K\"ahler moduli space is isomorphic to its own complex structure moduli space, hence $\hat{Z}$ is also realisable as a double elliptic fibration. Mirror symmetry thus predicts the existence of infinitely many special Lagrangian sub-manifolds in this Calabi-Yau and we construct these explicitly.

The plan of this paper is as follows. In Section \ref{sect:g2associatives}, we describe the class of compact $G_2$ orbifolds that we will consider, which are obtained as discrete quotients of $K3 \times T^3$. For a particular family of examples, we describe an associative sub-orbifold for every element of the lattice $E_8$. 
We also identify infinite numbers of special Lagrangian submanifolds in related Calabi-Yau orbifolds of Voisin-Borcea type which are expected to persist in the smooth crepant resolutions. In Section \ref{sect:IIAIIBF}, we give a description of this background in terms of the IIA string on an orientifold of a Calabi-Yau threefold of Voisin-Borcea type. By an application of mirror symmetry this is mapped to type IIB string theory, which has a straightforward lift to $F$-theory on a Calabi-Yau fourfold $Y$. By tracking the fate of the associative submanifolds of $M$ we started with, we show that they are in one-to-one correspondence with rigid divisors of $Y$. Finally, in Section~\ref{Sec:TCSorbif} we show how our orbifold example can be given the structure of a twisted connected sum, and find agreement with the work of \cite{Braun:2018fdp}. Interestingly, our $G_2$ orbifold example also admits another different decomposition as a twisted connected sum. 

\section{G2 orbifolds and Calibrated Cycles}\label{sect:g2associatives}

\subsection{General Considerations}

A 7-manifold $M^7$ whose holonomy group is the exceptional group $G_2$ may be characterised by the existence of a $G_2$-invariant 3-form $\varphi$. Since $G_2 \subset SO(7)$, $\varphi$ induces a Riemannian metric on $M^7$, $g_\varphi$. The holonomy group of this metric is a subgroup of $G_2$ if and only if $d\varphi =0$ and $d*_{\varphi}\varphi = 0$ i.e. closed and co-closed with respect to the induced metric. Such a metric is Ricci flat and, hence, provides a background model for $M$-theory. A model for $\varphi$ on any given tangent space can be written as follows. Choose a linear decomposition, $\R^7 = \R^4 \oplus \R^3$ together with orthonormal bases, $\omega_i^0$ and $dx_i$, for $\Lambda^2_+ (\R^4)$ and the 1-forms on $\R^3$. Then,
\begin{equation}
\begin{aligned}
\varphi_0  & =  \sum_i \omega_i^0 \wedge dx_i - dx_1 \wedge dx_2 \wedge dx_3 \:, \\
*\varphi_0 & = \tfrac16 \sum_i \omega^0_i \wedge \omega^0_i - \tfrac12 \sum_i \omega_i \wedge dx_j \wedge dx_k \epsilon^{ijk} \:.
\end{aligned}
\end{equation}

The first compact manifolds with holonomy $G_2$ were constructed by Joyce \cite{Joyce} via a gluing construction. The general idea is to first start with a manifold with a much smaller 
holonomy group $\subset G_2$ which admits a singular, discrete quotient; then  
provided that the quotients are suitably chosen and that one has suitable special holonomy model metrics that can be glued in around the excised singular regions, one can produce a simply connected 7-manifold with a $G_2$-structure which is close to $G_2$-holonomy (i.e. small torsion). Joyce's powerful perturbation theory argument then demonstrates that if the torsion is suitably well controlled, then there exist a nearby $G_2$-holonomy structure.
In most of Joyce's examples, the starting manifold is either a flat 7-torus $T^7$ or $X \times T^3$, with $X$ a hyper-K\"ahler K3 surface and the quotients are chosen so that the singularities may be removed by gluing in model metrics with either $SU(2)$ or $SU(3)$ holonomy. Even though these local models have holonomy strictly smaller than $G_2$, the fact that the construction produces a simply connected 7-manifold, implies that a closed and co-closed $G_2$-structure corresponds to a metric with holonomy group exactly $G_2$.

In this paper, our main interest will be in $G_2$ orbifolds $M$ of the form
\begin{equation}\label{eq:defM}
M = \left( X \times T^3 \right)/\Z_2 \times \Z_2   \:,
\end{equation}
where $X$ is a K3 surface admitting two simultaneous, isometric involutions such that $M$ has a well defined torsion free $G_2$ structure. In some cases, one can use Joyce's results to produce smooth 7-manifolds with $G_2$-holonomy, but, in general, from a physics point of view, $M$-theory on $M$ is physically consistent even in the presence of such orbifold singularities. Such singular $G_2$-orbifolds were discussed in the physics literature in \cite{Acharya:1998pm} and more recently in 
\cite{Reidegeld:2015bgp,2017arXiv170709325J}.

In terms of the hyper-K\"ahler structure $\omega_i$, $i =1,2,3$ on the K3 surface $X$, and the one-forms $dx_i$ on $T^3$, the $G_2$ structure on $X \times T^3$ can be written as 
\begin{equation}\label{eq:PHI3}
\begin{aligned}
 \varphi & = \sum_i \omega_i \wedge dx_i - dx_1 \wedge dx_2 \wedge dx_3 \:,\\
 *\varphi & = \tfrac16 \sum_i \omega_i \wedge \omega_i + \tfrac12 \sum_i \omega_i \wedge dx_j \wedge dx_k \epsilon^{ijk} \:.
\end{aligned}\hspace{1cm}
\end{equation}
We are then interested in a commuting pair of isometric involutions on $X \times T^3$ (with generators denoted by $\alpha$ and $\beta$) whose action on the forms is given by
\begin{equation}\label{eq:actiononHK}
 \begin{array}{ccccccc}
  & \omega_1 & \omega_2 & \omega_3 & dx_1 & dx_2 & dx_3 \\
\alpha & + & - & - & + & - & - \\
\beta & - & - & + & - & - & +
 \end{array}\, .
\end{equation}
Clearly, this preserves the $G_2$-structure which therefore descends to a torsion free $G_2$-structure on $M$. Note that, in any given complex structure on $X$, one out of the three involutions $\alpha$, $\beta$ or $\alpha\beta$ will act holomorphically with the other two acting anti-holomorphically. We will usually choose $\alpha$ to be holomorphic and $\beta$ to be anti-holomorphic.

In order to discuss K3 surfaces admitting involutions, it will be useful to briefly recall key facts about the moduli space of hyper-K\"ahler metrics on $X$. This is described by the global Torelli theorem, see e.g. \cite{barth1984compact, Aspinwall:1996mn} for a review. The second cohomology group of $X$ is rank 22 and given by $H^2(X,\Z)\otimes \R = \left(U^{\oplus 3} \oplus (-E_8)^2 \right)\otimes \R \equiv \R^{3,19}$ with the $(3,19)$ signature arising from the self and anti-self dual harmonic two forms on $X$ together with an inner form given by the cup product. In the K3 lattice $\left(U^{\oplus 3} \oplus (-E_8)^2 \right)$,  $U$ is the standard hyperbolic lattice of signature $(1,1)$ and $E_8$ is the root lattice of the corresponding Lie algebra. The space of hyper-K\"ahler metrics on $X$ is given by the Grassmanian of {\it space-like} three-planes in $\R^{3,19}$, up to discrete identifications. A choice of three orthonormal two-forms $\omega_i$ spanning such a three-plane yields the hyper-K\"ahler structure on $X$. This is locally equivalent to viewing the space of hyper-K\"ahler metrics on $X$ as the space of period integrals of the three $\omega_i$ over a basis of 2-cycles Poincar\'e dual to the 19 anti-self dual, timelike directions in $H^2(X,\R)$. A hyper-K\"ahler K3 surface $X$ will be singular if $H^2(X,\Z)$ contains at least one 2-sphere of self-intersection number minus two for which all three periods vanish. Such a 2-sphere has zero volume and the corresponding $X$ will have an orbifold singularity of $ADE$-type.

For any of the non-trivial elements in $\Z_2^\alpha \times \Z_2^\beta$, there exists a complex structure in which the K\"ahler form of $X$ is even, whilst the holomorphic two-form is odd. K3 surfaces which admit such involutions are rather special since a generic hyper-K\"ahler metric on $X$ will not have any symmetries at all. In fact,
such involutions have been classified in terms of a triple of invariants $(r,a,\delta)$ \cite{nikulin1976finite,0025-5726-14-1-A06,2004math......6536A}. Here $r$ is the rank of the sublattice $H^2_+(X,\Z)$ of $H^2(X,\Z)$ which is even under the involution,
$a$ is determined by its discriminant group $\Gamma \equiv H^2_+(X,\Z)^{ \ast}/H^2_+(X,\Z)=\Z_2^a$ and $\delta$ is $0$ if the bilinear form on $H^2(X,\Z)$ restricted to $\Gamma$ is integral and is $1$ otherwise (see \cite{0025-5726-14-1-A06} for some more background). 
Except for two special cases, the fixed point set of the involution is then given by the disjoint union of a curve of genus
$g = \tfrac12\left(22 -r -a \right)$ and $k = \tfrac12(r-a)$ rational curves. The two special cases are $(r,a,\delta)=(10,10,0)$, for which the involution acts freely, and $(r,a,\delta)=(10,8,0)$, for which the fixed point set consists of two disjoint elliptic curves.

If we consider K3 surfaces which admit both involutions $\alpha$ and $\beta$, $H^2(X,\Z)$ splits into four sublattices corresponding to the irreducible representations of $\Z_2^\alpha \times \Z_2^\beta$:
\begin{equation}
H^2(X,\Z) \supseteq \Lambda_{--} \oplus\Lambda_{-+} \oplus\Lambda_{+-} \oplus\Lambda_{++}  \:.
\end{equation}
Since we require $\varphi$ to descend to $M$ \eqref{eq:actiononHK},  such symmetric $K3$ surfaces have hyper-K\"ahler structures which are restricted to satisfy:
\begin{equation}\label{omegailatticesGEN}
\begin{aligned}
\omega_1 & \in \Lambda_{+-} \otimes \R  \\
\omega_2 & \in \Lambda_{--} \otimes \R  \\
\omega_3 & \in \Lambda_{-+} \otimes \R 
\end{aligned}
\end{equation} 
Hence, the three-plane spanned by $\omega_i$ has zero components along the subspace $\Lambda_{++}\otimes\R$.

Calibrated submanifolds (and orbifolds) are a particularly distinguished class of subspaces of special holonomy spaces \cite{harvey1982calibrated}. Physically, they are also known as ``supersymmetric cycles" \cite{Becker:1996ay} and are natural energy-minimising subspaces for branes to wrap. Associative submanifolds are three-dimensional submanifolds, $C$, in a $G_2$-holonomy space for which $\int_C \varphi = Vol(C)$ (up to a choice of orientation). This definition naturally extends to the orbifolds under consideration here. Given the form of $\varphi$ above, a natural class of associatives in $M$ arise either from holomorphic curves $\Sigma \subset X$ or special Lagrangian surfaces in $X$. The former are calibrated by the K\"ahler form, the latter by the real part of a holomorphic volume form. Clearly the two classes can be related by rotating the complex structures on $X$. The pairing in $\varphi$ between the hyper-K\"ahler forms $\omega_i$ and the $dx_i$ is $SO(3)$ invariant, so the product of a holomorphic curve calibrated by, say, $\omega_i$ and the coordinate circle $\subset T^3$ calibrated by $dx_i$ is an associative in $X\times T^3$. If, furthermore, $\Sigma \times S^1$ is mapped to itself by both $\alpha$ and $\beta$, then $C = (\Sigma \times S^1)/\Z_2^\alpha \times \Z_2^\beta$ will be an associative sub-orbifold in the $G_2$-orbifold $M$. As we will see, this setting allows for the construction of examples with infinitely many, homologically distinct associative 3-spheres in certain $G_2$-orbifolds.

\subsection{A model with an Infinity of Associative Cycles}\label{sect:mainexample}

Here, we will focus on a particular family of examples with a simple and explicit algebraic realisation for which we will be able to construct associatives and co-associatives (four-cycles calibrated by $*\varphi$) in $M$.

\subsubsection{Description of the Model}\label{sec:exampleWithInfAss}

In these examples, we realise $X$ as a family of elliptic K3 surfaces given as a complete intersection in a bundle of weighted projective planes $ \P^2_{321}$ over $\P^2$:
\begin{equation}\label{eq:k3algrealization}
\left\{\,\,\begin{aligned}
  y^2 & =  x^3 + x w^4 f_4(z) + w^6 g_6(z) \\
 \xi^2 & = z_1 z_2 
\end{aligned}\right. \:,
\end{equation}
where $[\xi:z_1:z_2]$ are homogeneous coordinates on $\P^2$, $[y:x:w]$ are homogeneous coordinates on $\P^2_{321}$ and $f_4,g_6$ are homogeneous polynomials in $z_1,z_2$ of the indicated degree. If we choose the coefficients of all monomials in $f_4,g_6$ to be real, this family admits two appropriate commuting involutions:
\begin{align}
\alpha: \hspace{.5cm} & \xi \rightarrow -\xi  \:,\\
\beta:\hspace{.5cm}& (y,x,w,z_1,z_2,\xi)  \rightarrow  (\bar{y},\bar{x},\bar{w},\bar{z_1},\bar{z_2},\bar{\xi})  \:.
\end{align}
If we associate $\omega_1 = J_X$ with the K\"ahler form and $\omega_2 + i\omega_3 = \Omega_X^{2,0}$ with the holomorphic two-form on the surface given by \eqref{eq:k3algrealization} and in the complex structure inherited from the ambient space, $\alpha$ and $\beta$ precisely act holomorphically and anti-holomorphically as demanded by \eqref{eq:actiononHK}. That is, $J_X\rightarrow J_X$ and $\Omega_X\rightarrow-\Omega_X$ under $\alpha$, while $J_X\rightarrow - J_X$ and $\Omega_X\rightarrow-\bar\Omega_X$ under $\beta$.

The fixed point set of $\alpha$ consists of the two elliptic curves at $z_1 = 0$ and $z_2 =0$, hence
it is the involution with invariants  $(r,a,\delta) = (10,8,0)$ in the language of \cite{nikulin1976finite,0025-5726-14-1-A06,2004math......6536A}. This means that it acts on the K3 lattice as 
\begin{equation}\label{alphaInvol}
\alpha \,\, \,\mbox{\Large\rotatebox[origin=c]{90}{$\curvearrowleft$}}\,\,\,
\begin{array}{rrrrr}
E_8 & E_8^\prime & U_1 & U_2 & U_3 \\
 \hline
E_8^\prime & E_8 & U_1 & -U_2 & -U_3 
\end{array}\, ,
\end{equation}
where we used a $^\prime$ to distinguish the two $E_8$ factors. So, in this example, the $\alpha$-invariant sublattice is:
\begin{equation}
H^2_+(X,\mathbb{Z})=(-E^+_8\oplus U_1)\:,
\end{equation}
where $E_8^+$ is the diagonal sublattice in $E_8 \oplus E_8^\prime$. The quotient surface $S=X/\alpha$ is a rational elliptic surface and $H^2_+(X,\mathbb{Z}) = H^2(S,\Z)$. Furthermore,  all points in this lattice are also points in $H^{1,1}(S,\R)$ (since $H^{2,0}(S)=0$), so they are dual to holomorphic cycles. Their lifts to $X$ are therefore also holomorphic.

The fixed points of $\beta$ depend on the specific choice of the polynomials $f$ and $g$: First of all notice that $\beta$ acts on the $S^2$ base  of the elliptic K3 by fixing one circle given by $\xi^2=z_1z_2$ with $[\xi:z_1:z_2]\in \R\P^2$. The zeroes of the  discriminant $\Delta(z_1,z_2)\equiv 4f^3-27g^2$ of the elliptic fibration are either real or in pairs of complex conjugates ($f$ and $g$ are polynomials with real coefficients). If we choose $f$ and $g$ such that no zeroes belong to the fixed circle, then the fixed point locus of $\beta$ in K3 is given by a pair of 2-tori. With this choice, $\beta$ is also of type $(r,a,\delta) = (10,8,0)$ and acts as
\begin{equation}\label{eq:actionz2oa2}
\beta \,\, \,\mbox{\Large\rotatebox[origin=c]{90}{$\curvearrowleft$}}\,\,\,
\begin{array}{rrrrr}
E_8 & E_8^\prime & U_1 & U_2 & U_3 \\
 \hline
-E_8^\prime & -E_8 & -U_1 & -U_2 & U_3
\end{array}\:.
\end{equation}
Note that, even though $\beta$ is acting anti-holomorphically with our choice of complex structure, there exists a different complex structure for which it's action is holomorphic.

The product $\alpha\beta$ is given by
\begin{equation}\label{eq:actionz2oa2ab}
\alpha \beta \,\, \,\mbox{\Large\rotatebox[origin=c]{90}{$\curvearrowleft$}}\,\,\,
\begin{array}{rrrrr}
E_8 & E_8^\prime & U_1 & U_2 & U_3 \\
 \hline
-E_8 & -E_8^\prime & -U_1 & U_2 & -U_3\:.    
\end{array}\:.
\end{equation}
One can further show that this involution is of type  $(r,a,\delta) = (2,0,0)$, whose fixed point set is a 2-sphere and a genus 10 curve, $\Sigma_{10}$.

\

The singular set of $M$ consists of eight disjoint $T^3$'s fixed by $\alpha$, another eight disjoint $T^3$'s fixed by $\beta$, four 3-spheres of the form $(S^2\times S^1_{x_2})/\alpha$, and four 3-orbifolds $(\Sigma_{10}\times S^1_{x_2})/{\alpha}$.
One expects that the $T^3$ singularities can be resolved using techniques from \cite{Joyce}, though, because the fixed points of $\alpha$ intersect those of $\beta$ in $X$, this will only be a partial desingularisation. 

\

\subsubsection{Calabi-Yau Examples}\label{Sec:VBmanf}

By considering only one involution instead of two, one can obtain examples which are products of Calabi-Yau threefolds and a circle. We briefly mention these here as they may be of independent interest.
For instance, $(X \times T^3)/\alpha = Z_{\alpha} \times S_{x_1}^1$ since $\alpha$ does not act on $x_1$. One can similarly define $Z_\beta$ and $Z_{\alpha\beta}$. Let us consider $Z_\alpha = (X \times T^2)/\alpha$. This is a Calabi-Yau orbifold with K\"ahler and holomorphic 3-form given by
\begin{equation}
\begin{aligned}
J_{Z_\alpha} &= \omega_1 + dx_2\wedge dx_3 \:, \\
\Omega_{Z_\alpha} &= (\omega_2 + i\omega_3)\wedge(dx_2+idx_3)  \:.
\end{aligned}\hspace{.5cm}
\end{equation}
The Calabi-Yau structures of $Z_{\beta}$ and $Z_{\alpha\beta}$ can be obtained by the obvious permutations. All three of these cases are examples of orbifolds considered by Borcea and Voisin \cite{borcea1997k3,voisin1993miroirs}, who showed, using \cite{bridgeland2001mckay} that they admit crepant resolutions which are smooth Calabi-Yau threefolds $\tilde{Z}_{\sigma}$, where $\sigma$ is any such involution of $X$. In particular, if $\sigma$ is of type $(r,a,\delta)$ then the non-trivial Hodge numbers of $\tilde{Z}_{\sigma}$ are
\begin{equation}
h^{1,1} = 5 + 3r -2a\:, \hspace{1cm} h^{2,1} = 65 - 3r - 2a\:.
\end{equation}
So, in our particular examples, $(h^{1,1}(\tilde{Z}_{\sigma}),h^{2,1}(\tilde{Z}_{\sigma})) = (19,19), (19,19), (11,59)$ for $\sigma = \alpha, \beta, \alpha\beta$ respectively.
The Calabi-Yau threefolds in the $(19,19)$ cases were first considered in \cite{gukov2003duality} and can be regarded as a fibre product of two copies of the rational elliptic surface over a common $\P^1$. We will later construct an infinite number of special Lagrangian 3-cycles in $Z_\alpha$ and argue that they persist in the resolution $\tilde{Z}_\alpha$. $\tilde{Z}_\beta$ similarly gives an equivalent threefold.

\subsubsection{Constructing Associatives and Co-associatives in $M$}\label{sec:Associatives}

We now show that the $G_2$-orbifold $M$ possesses infinitely many calibrated sub-orbifolds. The basic underlying reason for this is that $X$ is an elliptically fibred $K3$ surface with infinitely many sections. The sections of a compact, elliptically fibred complex surface, $Y$, form an abelian group, called the Mordell-Weil group $MW(Y)$ of $Y$, and the zero section is the identity element of this group. In our case, $X/\alpha = S$ is a rational elliptic surface and, for generic choice of parameters in \eqref{eq:k3algrealization} $S$ has twelve singular fibres. In this generic case, $MW(S)\cong E_8^+$ \cite{shioda1990mordell}, so, for every element in the root lattice of $E_8$ there is a holomorphic 2-sphere in $S$. This lifts to a holomorphic 2-sphere in $X$, preserved by $\alpha$. We will now describe the homology classes of these 2-spheres.

The existence of the two involutions $\alpha$ and $\beta$ (or, equivalently, $X$ being a member of the algebraic family \eqref{eq:k3algrealization}), constrains the hyper K\"ahler structure consistently with \eqref{omegailatticesGEN}, i.e. 
\begin{equation}\label{omegailattices}
\begin{aligned}
\omega_1 & \in \left(  -E^+_8  \oplus U_1\right) \otimes \R  \\
\omega_2 & \in U_2  \otimes \R  \\
\omega_3 & \in \left( -E_8^- \oplus U_3\right) \otimes \R 
\end{aligned}
\end{equation}
where $-E_8^\pm$ denote the diagonal sublattices of $E_8^+ \oplus E_8^-$, which contain elements of the form $\gamma \pm \gamma^\prime$ with $\gamma \in -E_8$ and $\gamma^\prime \in -E_8^\prime$ such that $\gamma = \gamma'$ under the isomorphism between $E_8^+$ and $E_8^-$ fixed by $\alpha$. 
Denoting the generators of $U_1$ by $e_1$ and $e^1$, we may identify the class of the zero section of the elliptic fibration on $X$ with $\sigma_0 = e_1-e^1$ and the class of a generic fibre with $F = e^1$. 

For every lattice element $\gamma$ in $-E_8$ with $\gamma^2 = -2n$, we may now consider the class \cite{Donagi:1996yf}
\begin{equation}\label{eq:sigmagammas}
 \sigma_\gamma = \sigma_0 + 2 n F + \gamma + \gamma^\prime \, ,
\end{equation}
which satisfies $\sigma_\gamma^2 = -2$ and $F\cdot \sigma_\gamma = 1$. Since $\sigma_\gamma$ is an element of the Picard lattice with self-intersection -2, there exists a rational curve (i.e. an $S^2$) $\Sigma_\gamma \subset X$ in this homology class. The property $\sigma_\gamma \cdot F =1$ furthermore 
implies that these $S^2$'s are precisely the sections of the elliptic fibration of $X$. Notice that since $\alpha$ acts holomorphically on each $\Sigma_{\gamma}$, the action will have two fixed points.

As the second involution $\beta$ is anti-holomorphic, it acts on the sections $\Sigma_\gamma$ by sending every point on $\Sigma_\gamma$ to its complex conjugate point in $X$:
\begin{equation}
\beta: ( y(z),x(z),w(z),z_1,z_2,\xi) \rightarrow (\bar{y(z)},\bar{x(z)},\bar{w(z)},\bar{z_1},\bar{z_2},\bar{\xi}) \, .
\end{equation}
We now show that $\Sigma_\gamma$, seen as a submanifold of $S$, is in fact mapped back to itself by $\beta$. This follows from the two facts that $\beta$ maps the class of $\Sigma_\gamma$ to minus itself and that each section has a unique holomorphic representative (as shown in \cite{Donagi:1996yf}). Preserving  $\Sigma_\gamma$ as a submanifold is equivalent to showing that
\begin{equation}\label{eq:holhol}
(\bar{y(z)},\bar{x(z)},\bar{w(z)},\bar{z}_1,\bar{z}_2) = (y(\bar{z}),x(\bar{z}),w(\bar{z}),\bar{z}_1,\bar{z}_2) \, .
\end{equation}
In fact, $z\rightarrow \bar{z}$ simply reverses the orientation on the $\mathbb{P}^1$ base\footnote{This is responsible for $\beta:\sigma_\gamma \rightarrow -\sigma_\gamma$.}; then \eqref{eq:holhol} means that we end up at a different point on the same submanifold $\Sigma_\gamma$. To prove our assertion, assume that \eqref{eq:holhol} does not hold, i.e. that
\begin{equation}
 (\bar{y(z)},\bar{x(z)},\bar{w(z)},\bar{z}_1,\bar{z}_2) = (y'(\bar{z}),x'(\bar{z}),w'(\bar{z}),\bar{z}_1,\bar{z}_2) \, ,
\end{equation}
for some functions $(y',x',w')\neq (y,x,w)$. We can now reverse the orientation on the base again by mapping $z\rightarrow \bar{z}$ to find a holomorphic section described by the functions $y',x',w'$. As each section is uniquely specified by its homology class, this section cannot possibly be in the class $\sigma_\gamma$, and we find a contradiction with the fact that that $\beta:\sigma_\gamma \rightarrow -\sigma_\gamma$. 

Since $\Sigma_\gamma$ is preserved by both $\alpha$ and $\beta$, we have thus shown that
\begin{equation}
C_\gamma = \left( \Sigma_\gamma \times S^1_{x_1}\right) /\Z_2^\alpha \times \Z_2^\beta
\end{equation}
is an associative sub-orbifold in $M$. The topology of $C_\gamma$ is in fact that of $S^3$. To see this, we first note that, since $\alpha$ acts holomorphically on $\Sigma_{\gamma}$, $\Sigma_{\gamma}/\alpha$ is also a 2-sphere. Now $\beta$ acts on this 2-sphere as complex conjugation, fixing an $S^1 \subset \Sigma_{\gamma}/\alpha$ and two points on $S^1_{x_1}$. But since codimension two fixed points do not contribute to the fundamental group of $C_\gamma$ and the action of $\beta$ projects out the 1-cycles in $S^1_{x_1}$, the fundamental group of $C_\gamma$ is trivial.
We can also describe $C_\gamma$ as an $S^2 = \Sigma_{\gamma}/\alpha$ sitting over an interval $S^1_{x_1}/\beta$ which degenerates at the ends.

Since the $C_\gamma$'s are all 3-spheres, we expect that they will contribute to the non-perturbative superpotential for compactifications of $M$-theory on $M$ \cite{Harvey:1999as}. 

\vspace{2mm}
As one can see, the example is symmetric under the exchange of $1\leftrightarrow 3$ and $\alpha\leftrightarrow \beta$. There must then be another set of associatives. In fact, consider a curve $\tilde{\Sigma}_\gamma$ dual to 
\begin{equation}
\tilde{\sigma}_\gamma =  \tilde{\sigma}_0 + 2n \tilde{F} + \gamma_+ - \gamma_-\, ,
\end{equation}
where $\tilde{\sigma}_0 = e_3-e^3$ and $\tilde{F} = e^3$. In the complex structure where $\omega_3$ is the K\"ahler structure, these are again sections of an elliptic fibration, but now one with base dual to $\tilde{\sigma}_0$ and fibre dual to $\tilde{F}$. These curves are calibrated by $\omega_3$. There is again a unique holomorphic representative for $\tilde{\Sigma}_\gamma$, which is preserved (by repeating the same argument as above) by both of the involutions $\alpha$ and $\beta$. The new set of associative cycles is given by
\begin{equation}
\tilde{C}_\gamma = \left( \tilde\Sigma_\gamma \times S^1_{x_3}\right) /\Z_2^\alpha \times \Z_2^\beta \:.
\end{equation}
Finally there is another associative with the topology of a three-sphere:  the two-sphere $\hat{\Sigma}$ in the class $e_2-e^2$ is a section of an elliptic fibration with fibre $e^2$ and it is calibrated by $\omega_2$. By the same consideration done above, the new associative is
\begin{equation}
\hat{C} = \left( \hat\Sigma \times S^1_{x_2}\right) /\Z_2^\alpha \times \Z_2^\beta \:.
\end{equation}

By analogous considerations one can also construct co-associative 4-cycles, i.e. 4-cycles calibrated by~$\Psi$:
\begin{eqnarray}
B_\gamma &=&   \left( \Sigma_\gamma \times  S^1_{x_2} \times S^1_{x_3}\right)  /\Z_2^\alpha \times \Z_2^\beta \:, \\
\tilde{B}_\gamma &=&  \left( \tilde{\Sigma}_\gamma \times  S^1_{x_1} \times S^1_{x_2}\right)  /\Z_2^\alpha \times \Z_2^\beta \:, \\
\hat{B} &=&   \left( \hat{\Sigma} \times  S^1_{x_1} \times S^1_{x_3}\right)  /\Z_2^\alpha \times \Z_2^\beta \:.
\end{eqnarray}

The topology of all of these 4-cycles is in fact that of an anti-holomorphic quotient of a rational elliptic surface with orbifold singularities.

\subsubsection{Calibrated Submanifolds in Borcea-Voisin Threefolds}

In the Calabi-Yau case we have the following examples. Recall the smooth Calabi-Yau threefold $\tilde{Z}_\alpha$ obtained by desingularising $Z_{\alpha} = (X\times T^2)/\alpha$. The 2-spheres $\Sigma_\gamma/\alpha \subset Z_{\alpha}$ are all calibrated by $\omega_1$. After the crepant resolution, these all become smooth rational curves in the Schoen threefold $\tilde{Z}_\alpha$. Furthermore, $(\Sigma_\gamma \times T^2)/\alpha \subset Z_\alpha$ are all holomorphic divisors in $Z_{\alpha}$. These persist as smooth divisors $D_\gamma$ in 
$\tilde{Z}_\alpha$. In fact, one can see that, as complex surfaces, each of the $D_\gamma$ is itself a rational elliptic surface. This reproduces a result in \cite{hosono1997mirror}.

Perhaps more interestingly, one can construct infinitely many special Lagrangian 3-spheres in $Z_\alpha$. These are simply given by $L_\gamma = (\tilde{\Sigma}_\gamma \times S^1_{x_3})/\alpha$. These are calibrated by $\omega_3\wedge dx_3$, which is the imaginary part of the holomorphic 3-form. When we resolve the orbifold singularities of $Z_{\alpha}$ to produce the manifold $\tilde{Z}_\alpha$, each of these mildly singular $S^3$'s $L_\gamma$ becomes a smooth $S^3$ in $\tilde{Z}_\alpha$ and it is very natural to conjecture that they are all special Lagrangians in $\tilde{Z}_\alpha$. This is supported by the fact that the crepant resolution changes only the K\"ahler class of the orbifold and not its complex structure. This is also expected from mirror symmetry considerations.

\section{Description in IIA, IIB and $F$-theory}\label{sect:IIAIIBF}

In the following, we will consider duality maps from $M$-theory to type IIA string theory, and then consider the type IIB mirror and its $F$-theory lift. 

\subsection{Type IIA}\label{sec:TypeIIA}

Compactification of $M$-theory on $M$ has a description in terms of type IIA string theory by doing a circle reduction along one of $S^1_{x_i}$ ($i=1,2,3$). Choosing different circles in principle gives different (dual) answers in type IIA.
Here we choose to reduce along $x_3$. In this description, $\beta$ creates the Calabi-Yau orbifold
\begin{equation}
Z_\beta = ( X \times S^1_{x_1} \times S^1_{x_2} )/\beta   \:,
\end{equation}
which is a singular limit of the Schoen Calabi-Yau threefold with Hodge numbers $(19,19)$~\cite{Donagi:2008xy}, as we have described in Section~\ref{Sec:VBmanf}. The complex structure on the $K3$ surface $X$ is chosen such that $\omega_3$ is the K\"ahler form on $X$ and $\Omega_X=\omega_2 + i \omega_1$ is the holomorphic $(2,0)$ form on $X$. We have also introduced a holomorphic coordinate $z=x_2+i\,x_1$ on $T^2=S^1_{x_1} \times S^1_{x_2}$.

The  K\"ahler form and the holomorphic three-form on $Z$ are given by
\begin{equation}
\begin{aligned}
J_Z & = \omega_3 + dx_2 \wedge dx_1  \:, \\
\Omega_Z & =\Omega_X \wedge dz \:.
\end{aligned} 
\end{equation}
The second involution $\alpha$ is now an (anti-holomorphic) orientifold involution, which consequently acts with a sign on the $M$-theory circle. 

In type IIA CY compactifications, the orientifold involution $\iota$ must be anti-holomorphic and isometric, i.e. the K\"ahler form must transform as $\iota^\ast J_Z = -J_Z$ and the holomorphic $(3,0)$-form as $\iota^\ast \Omega_Z^{3,0}=e^{i\theta}\bar{\Omega}_Z^{0,3}$,  where $\iota^*$ is the pullback of $\iota$ and $e^{i\theta}$ is a constant phase that we will set equal to~1.
Such an involution generically admits codimension-3 fixed point loci, the O6-planes. These loci are special Lagrangian three-cycles in the CY, that we call $\Pi_{O6}$; in fact, as a consequence of the action of $\iota$ on $J_Z$ and $\Omega_Z^{3,0}$ and the fact that the points of $\Pi_{O6}$ are fixed, one has
\begin{equation}
J_Z|_{\Pi_{O6}} =0 \,, \qquad \qquad \mbox{Im}\Omega^{3,0}_Z|_{\Pi_{O6}}=0\:.
\end{equation}
The fixed point locus is given by the fixed points of $\alpha$ and the fixed points of $\alpha\beta$. The O6-plane three-cycles are a product of a fixed curve in K3 and one of the torus circles, modded by the orbifold involution $\beta$. 
On K3, the fixed point locus of $\alpha$, a $(10,8,0)$ involution, is given by two 2-tori $T^2_{1,2}$ in the $U_1$ block. On the 2-torus the involution acts fixing two points on the circle $S^1_{x_2}$. Hence the fixed three-cycles are two copies of
\begin{equation}\label{FixedAlphaIIA}
(T^2_{1}\times S^1_{x_1})/\beta,  \qquad (T^2_{2}\times S^1_{x_1})/\beta  .
\end{equation}
The fixed point locus of $\alpha\beta$, a $(2,0,0)$ involution, is given by a disjoint union of a 2-sphere  $\eta_{0}$ and a genus $g=10$ curve $\eta_{10}$ in the $U_2$ block \cite{nikulin1976finite,0025-5726-14-1-A06,2004math......6536A}.
Since both must be given by a linear combination of $e_2,e^2$ (a fixed curve must be in an even class and the only even cycles with respect to $\alpha\beta$ are in the $U_2$ block; see \eqref{eq:actionz2oa2ab}), it is easy to work out their homology classes: 
$[\eta_{0}]= e_2-e^2$ and $[\eta_{10}] = 3 e_2 + 3 e^2$. The involution $\alpha\beta$  also fixes two points on $S^1_{x_1}$. The corresponding O6-planes are  then two copies of 
\begin{equation}\label{FixedAlphaBetaIIA}
 (\eta_{0}\times S^1_{x_2})/\beta, \qquad  (\eta_{10}\times S^1_{x_2})/\beta .
 \end{equation}
All the O6-plane three-cycles are Lagrangian submanifolds of $Z$, calibrated by Re$(\Omega_Z^{3,0})$.

The associative and the coassociative cycles now give rise to the  special Lagrangian submanifolds
\begin{equation}\label{eq:sLagsinIIA}
\begin{aligned}
C_\gamma^{IIA} & = \left(\Sigma_\gamma \times S^1_{x_1}\right)/ \beta \:, \\
\hat{C}^{IIA} & = \left(\hat{\Sigma} \times S^1_{x_2}\right)/ \beta \:, \\
B_\gamma^{IIA} & = \left(\Sigma_\gamma \times S^1_{x_2}\right)/ \beta \:, \\
\hat{B}^{IIA} & = \left(\hat{\Sigma} \times S^1_{x_1}\right)/ \beta \:,
\end{aligned}
\end{equation}
the holomorphic curves
\begin{equation}
\begin{aligned}
\tilde{C}_\gamma^{IIA} & = \left(\tilde{\Sigma}_\gamma \right)/ \beta 
\end{aligned}
\end{equation}
and the divisors
\begin{equation}
\begin{aligned}
\tilde{B}_\gamma^{IIA} & = \left(\tilde{\Sigma}_\gamma \times S^1_{x_1}\times S^1_{x_2}\right)/ \beta\, .
\end{aligned}
\end{equation}

The cycles $C_\gamma^{IIA}$ and $\hat{C}^{IIA}$ are calibrated by Re$(\Omega_Z^{3,0})$;   the cycles $B_\gamma^{IIA}$ and $\hat{B}^{IIA}$ are calibrated by Im$(\Omega_Z^{3,0})$; the cycles
$\tilde{B}_\gamma^{IIA}$ calibrated by $J_Z\wedge J_Z$ and the cycles $\tilde{B}_\gamma^{IIA}$ are calibrated by $J_Z$. The cycles $C_\gamma^{IIA}$, $\hat{C}^{IIA}$, $B_\gamma^{IIA}$ and $\hat{B}^{IIA}$ have the topology of three-spheres; the first two are even with respect to $\alpha$, while the last two are odd. The divisors $\tilde{B}_\gamma^{IIA}$ have the topology of $\P^1\times\P^1$, where the last factor is the punctured pillow. 

Since the O6-planes are calibrated by Re$(\Omega_Z^{3,0})$, only the special Lagrangian submanifolds $C_\gamma^{IIA}$ and $\hat{C}^{IIA}$ can be wrapped by BPS D2-instantons. This is consistent with the $M$-theory lift: M2-branes were wrapping the associative three-cycles $C_\gamma$, $\hat{C}$ and $\tilde C_\gamma$. The first two do not wrap the $M$-theory circle $S^1_{x_3}$ and give rise to D2-instantons in type IIA, while the last ones reduce to strings wrapping the curves $\tilde C_\gamma^{IIA}$, i.e. they are worldsheet instantons.

Note that in our chosen complex structure of $X$, the sections $\tilde\Sigma_\gamma$ are algebraic sections of the elliptic fibration of $X$, but the $\Sigma_\gamma$ no longer have an algebraic representation. Nonetheless, they can be depicted rather explicitly in terms of string junctions. We give a brief review of this picture, including an example of such a string junction in appendix \ref{app:StringJunctions}.

\subsection{Type IIB}\label{sect:IIB}

To get to type IIB, we need to study the mirror map of $Z_\beta$, which is composed of the mirror map on $X$ together with a T-duality along either $S^1_{x_1}$ or $S^1_{x_2}$. We choose to T-dualise along $x_2$ (in order to have D3/D7-branes). On $X$, the mirror map simply becomes the hyper-K\"ahler rotation which acts as \cite{dolgachev1996mirror,Aspinwall:1994rg,gross1998special}
\begin{align}
{\rm Im}(\Omega^{(2,0)}_{X})&\rightarrow\omega_{X^\vee}\\
\omega_{X}&\rightarrow-\,{\rm Im}(\Omega^{(2,0)}_{X^\vee})\\
{\rm Re}(\Omega^{(2,0)}_{X})&\rightarrow {\rm Re}(\Omega^{(2,0)}_{X^\vee})\:.
\end{align}
Hence\footnote{This mirror map exploits the elliptic fibration with fibre class $e^2$ as the SYZ fibration on $X$.}
\begin{equation}\label{IIBJOmega}
J_{X^\vee}=\omega_1 \qquad \mbox{and} \qquad \Omega_{X^\vee}=\omega_2-i\,\omega_3\, .
\end{equation}

The three-fold $Z^\vee_\beta$ is again an orbifold limit of the Schoen Calabi-Yau threefold, which is to be expected as this Calabi-Yau threefold is self-mirror \cite{hosono1997mirror}. The new orbifold involution $\beta^\vee$ inverts the sign of the torus coordinates, $(x_1,x_2)\mapsto (-x_1,-x_2)$, and it now acts as \cite{gross2000large}:
\begin{equation}\label{eq:actionz2ob2}
\beta^\vee \,\, \,\mbox{\Large\rotatebox[origin=c]{90}{$\curvearrowleft$}}\,\,\,
\begin{array}{rrrrr}
E_8^+ & E_8^- & U_1 & U_2 & U_3 \\
 \hline
E_8^- & E_8^+ & U_1 & -U_2 & -U_3
\end{array}
\end{equation}
on the K3 lattice.
This is consistent with the fact that the Picard lattice must be even, while the transcendental lattice must be odd (as one can check by using \eqref{IIBJOmega} and \eqref{omegailattices}).

The orientifold involution $\alpha$ acts as before on the K3 lattice (see \eqref{alphaInvol}) while sending $(x_1,x_2)\mapsto (x_1,x_2)$. The composition $\alpha \beta^\vee$ acts only on the $T^2$ as $(x_1,x_2)\mapsto (-x_1,-x_2)$.
The fixed point locus of this involution is given by six components, i.e. we have six O7 orientifold planes. There are four copies of $K3/\beta^\vee$, that have the topology of a rational elliptic surface, and two copies of $T^2\times \left(S^1_{x_1}\times S^1_{x_2}\right)/\alpha\beta^\vee$ with the topology of $T^2\times \mathbb{P}^1$.\footnote{The fixed point set of $\alpha$ is given by the fixed point locus of the $(10,8,0)$ Nikulin involution, that are two tori in the $U_1$-block. The fixed point locus of $\alpha\beta^\vee$ is given by four points in the $T^2_{x_1,x_2}$, modded by $\beta^\vee$.} 

One may explicitly describe the orbifold $Z^\vee_\beta=\left( X \times S^1_{x_1} \times S^1_{x_2}\right)/ \beta^\vee $ as the algebraic threefold
\begin{equation}\label{OrbifAlgeb}
\left\{\,\,\begin{aligned}
 y^2 & =  x^3 + x w^4 f_4(z) + w^6 g_6(z) \\
 \xi^2 & = z_1 z_2 \prod_{i=1}^4 (u_1 - \alpha_i u_2)
\end{aligned} \right.\:,
\end{equation}
where $[u_1:u_2]$ are homogeneous coordinates on another $\P^1$. The orientifold involution is simply $\xi \rightarrow -\xi$ and the fixed point locus $\xi=0$ splits manifestly in the six components mentioned above.
On $Z^\vee_\beta$, we can write 
\begin{equation}
\begin{aligned}
\Omega_{Z^{\vee}}^{3,0} & = \left( \omega_2 - i \omega_3 \right) \wedge \left(dx_2 - i \, dx_1\right) \:, \\
J_{Z^\vee} & = \omega_1 + dx_1 \wedge dx_2 \:.
\end{aligned}
\end{equation}

\begin{figure}[h!]
\begin{center}
  \includegraphics[height=7cm]{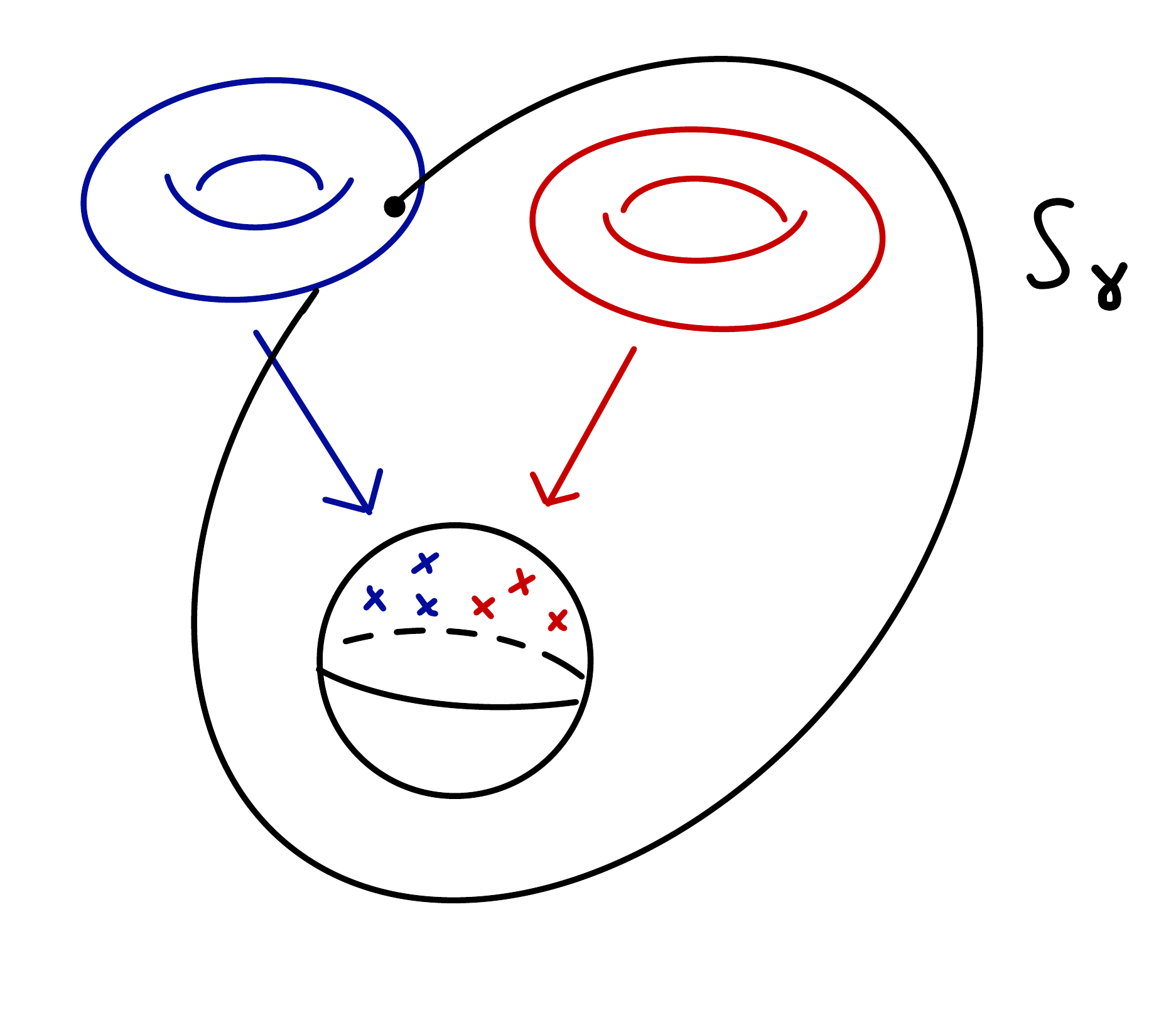}
 \caption{\label{fig:Schoen} The Schoen as a double elliptic fibration. The blue points on the base represent the degeneration loci of one elliptic fibration. The red dots represent the degeneration loci of the other elliptic fibration. Picking a section of the blue elliptic fibration leaves a divisor $S_\gamma$.}
\end{center}
\end{figure}

The special Lagrangian submanifolds $C_\gamma^{IIA}$ become the divisors
\begin{equation}
S_\gamma^{IIB} = \left(\Sigma_\gamma \times S^1_{x_1} \times S^1_{x_2}\right)/\mathbb{Z}_2^{\beta^\vee} \, 
\end{equation}
after the mirror map. They are calibrated by $J_{Z^\vee} \wedge J_{Z^\vee}$. In the Schoen Calabi-Yau threefold, which is a fibre-product of two rational elliptic surfaces, they appear as sections of the elliptic fibration of one of the two surfaces, so that they have the topology of a rational elliptic surface. This is depicted in Figure \ref{fig:Schoen}. 
In the algebraic realization they are points on the elliptic curve described by the first equation in \eqref{OrbifAlgeb}. One example is given by $w=0$.

These sections survive under deformations of the Schoen to a smooth geometry, where the second equation in \eqref{OrbifAlgeb} becomes $\xi^2=h_{2,4}(z,u)$ ($h_{2,4}$ is a generic polynomial of degree $(2,4)$).
In the smooth deformed CY threefold, the (orientifold) invariant divisors $S_\gamma^{IIB}$ get the topology of a smooth rational elliptic surface. This means that they are rigid holomorphic four-cycles with $h^{1,0}=0$, i.e. they satisfy the constraints sufficient to contribute to the superpotential.
In the dual $M$-theory setup, the deformation should correspond to a smoothing of the orbifold singularities in the $G_2$ holonomy manifold $M$ of Section \ref{sect:g2associatives}. The associatives $C_\gamma$ corresponding to $S_\gamma^{IIB}$ should be mapped to associative three-cycles in the smooth $G_2$-holonomy manifold.

\subsection{$F$-theory}

Type IIB string theory is the weakly coupled limit of $F$-theory. Given $F$-theory compactified on an elliptic fibration $Y$ over a base space $B_3$, the weak coupling limit can be worked out systematically \cite{Sen:1997gv}: one obtains perturbative type IIB on a Calabi-Yau threefold that is a double cover of $B_3$. In practice, let us consider the elliptically fibred four-fold $Y$ described by the equation 
\begin{equation}
 Y^2 = X^3+F(Z) X W^4 + G(Z) W^6
\end{equation}
where $F,G$ are sections respectively of $\bar{K}_{B_3}^{\otimes 4},\bar{K}_{B_3}^{\otimes 6}$, and $Z$ are coordinates on the base space $B_3$. The type IIB Calabi-Yau threefold is constructed by  adding a new coordinate $\xi$ that is a section of the anti-canonical bundle of $B_3$ and a new equation
\begin{equation}
\label{eq:doubble}
\xi^2=h( Z)\:.
\end{equation}
The orientifold involution is $\xi\mapsto -\xi$ and the fixed point set, i.e. the O7-plane locus, is given by the equation $h(Z)=0$.

When $F,G$ take the special form $F(Z)= c_F\,h^2(Z)$ and $G(Z)=c_G\,h^3(Z)$ ($c_F,c_G\in\mathbb{C}$), the elliptic fourfold has a $D_4$ singularity over the locus $h(Z)=0$ in the base. In the weak coupling limit, this corresponds to cancelling the D7-tadpole of the O7-plane by placing four D7-branes (plus their images) on top of the O7-plane itself \cite{Sen:1997gv}.

We can apply this considerations to the case under study in this paper. The base manifold $B_3$ is the quotient of the type IIB threefold $Z^\vee_\beta$ of Section~\ref{sect:IIB}, i.e. it is $S_{x,y,w;z}\times \mathbb{P}_{u}^1$ (where $S_{x,y,w;z}$ is a rational elliptic surface). Let us now look at the 
equations \eqref{OrbifAlgeb} defining the CY threefold in type IIB: the second equation is exactly of the form \eqref{eq:doubble}, while the first one defines the base manifold $B_3$. So the elliptic $F$-theory fourfold is given by
\begin{equation}\label{eq:DGWFth4fold}
\left\{\begin{array}{l}
Y^2=X^3+c_F\,h^2_{2,4}(z,u)X\,W^4 + c_G h^3_{2,4}(z,u) W^6 \\
y^2=x^3+f_4(z)\,x\,w^4+g_6(z)w^6 
\end{array}\right. \:,
\end{equation}
in the toric ambient space with weight system
\begin{equation}
\begin{tabular}{ c c c c c c c c cc| c  |  c | c }
  $X$ & $Y$ & $W$ & $x$ & $y$ & $w$ & $z_1$ & $z_2$ & $u_1$ & $u_2$ & Sum of degrees & EQ & eq \\\hline
  2 & 3 & 1 &  0 & 0 & 0 & 0 & 0 & 0 & 0 & 6 & 6 & 0\\
  0 & 0 & 0 &  2 & 3 & 1 & 0 & 0 & 0 & 0 & 6 & 0 & 6\\
  2 & 3 & 0 &  2 & 3 & 0 & 1 & 1 & 0 & 0  & 12 & 6 & 6\\
  4 & 6 & 0 &  0 & 0 & 0 & 0 & 0 & 1 & 1  & 12 & 12 & 0
\end{tabular}\, .
\end{equation}
If we now take 
$$h(z,u)=z_1 z_2 \prod_i (u_1 - \alpha_i u_2) \qquad (\alpha_i\in \mathbb{C}),$$
we obtain a fourfold with six $D_4$ singularities intersecting over curves in the base manifold $B_3$, that is the $F$-theory lift on type IIB on the orbifold $Z^\vee_\beta$ with D7-brane tadpole cancelled locally.
This fourfold is manifestly the Donagi-Grassi-Witten fourfold studied in \cite{Donagi:1996yf}, with a specific limit in the complex structure moduli space.\footnote{
This fourfold has the same $F$-theory limit of the orbifold 
$
Y = \left( Z_\beta^\vee \times E \right)/\Z_2^{\alpha\beta} ,
$
where $\alpha\beta$ also acts on the torus $E$ by $z \rightarrow -z$ \cite{Braun:2009wh}. }

Let us see how the interesting  divisors $S_\gamma^{IIB}$ lift to $F$-theory. These are invariant divisors under the involution $\xi\mapsto-\xi$. They project to the base $B_3$ to divisors $D^B_\gamma=S_\gamma^{IIB}/\alpha=\mathbb{P}^1_\gamma \times \mathbb{P}^1_{u}$, where $\mathbb{P}^1_\gamma$ are sections of the rational elliptic fibration $S_{x,y,w;z}$. They are lifted in $F$-theory to elliptic fibrations $D_\gamma^{F}$ over such divisors in the base.

As we said, the orbifold $Y$ is a singular limit of the Calabi-Yau fourfold considered in~\cite{Donagi:1996yf}, which is a generic elliptic fourfold with base $\P^1 \times S$. The divisors $D_\gamma^{F}$ that we found by dualities are precisely the rigid divisors which were argued in \cite{Donagi:1996yf} to give rise to a non-perturbative superpotential from Euclidean $D3$-brane instantons. 
In this paper we have shown that  these are naturally mapped to the associative three-cycles in $M$ we have found above under a chain of dualities, and which likewise give rise to a non-perturbative superpotential in the effective 4D $\mathcal{N}=1$ theory.

\section{Twisted Connected Sums}\label{Sec:TCSorbif}

$G_2$ manifolds are more difficult to construct and analyse than Calabi-Yau manifolds because complex algebraic geometry is not generally applicable to them. However, the thus far known constructions of compact
$G_2$ manifolds are all based on gluing constructions with algebraic building blocks \cite{Joyce,MR2024648,Corti:2012kd,MR3109862}.  The largest class of examples arise as twisted connected sums (TCS) \cite{MR2024648,Corti:2012kd,MR3109862}. This works as follows. An asymptotically cylindrical Calabi-Yau threefold $V$ is a non-compact complete Riemannian manifold with holonomy group $SU(3)$ which asymptotes to the Riemannian product of a K3 surface $\mathcal{S}$ and a cylinder $\R_+ \times S^1_v$ at a sufficiently rapid rate. Such threefolds can be constructed algebraically from compact K3-fibred algebraic threefolds $F$ with $c_1(F) = [\mathcal{S}]$ by excising a generic fibre, $V = F \setminus \mathcal{S}$ and an extension of Yau's theorem to this case establishes the existence of the Calabi-Yau metric \cite{MR3109862}.
For an appropriately chosen pair $(V_+, V_- )$ of asymptotically cylindrical Calabi-Yau threefolds, one forms a pair of 7-manifolds $M_{\pm} = V_\pm \times S^1_{w\pm}$ by taking their products with a circle. To produce a simply connected compact 7-manifold, one glues $M_+$ to $M_-$ along their asymptotic cylindrical regions, $\mathcal{S}_+ \times S^1_{v+} \times \R^+ \times S^1_{w+}$ and $\mathcal{S}_- \times S^1_{v-} \times \R^+ \times S^1_{w-}$. The gluing requires matching $\mathcal{S}_+$ to $\mathcal{S}_-$ with a hyper-K\"ahler rotation and exchanging the asymptotic circles in $V_{\pm}$, $S^1_{v\pm}$, with the external circles in $M_{\mp}$, $S^1_{w\mp}$. 

The fibrations by $\mathcal{S}_\pm$ give rise to a co-associative fibration with base $S^3$ (in the adiabatic limit) on $M$. The base $S^3$ is the result of gluing the two semi-infinite solid 2-tori $S^1_{v+} \times S^1_{w+} \times \mathbb{R}^+$ and $S^1_{v-} \times S^1_{w-} \times \mathbb{R}^+$ in the asymptotic regions.

\subsection{Realization of $M$ as a Twisted Connected Sum}
In our particular example, we have a compact $G_2$-orbifold, $M$.
As we will explain, $V_+$ is actually a locally flat orbifold, given by 
\begin{equation}\label{eq:Vplus}
V_+ = ( S^1_{\varepsilon_i} \times P_+ \times S^1_{x_1} \times S^1_{x_2} \times S^1_{x_3})/\Z_2^\alpha \times \Z_2^\beta \:,
\end{equation}
where  $S^1_{\varepsilon_i}$ is in $E$, the elliptic curve in the original K3-surface $X$, and $P_+\equiv S_{\zeta_i}^1\times\R^+$ is half  the $\P^1$ base of the elliptic fibration of $X$ in which $\beta$ was acting holomorphically. Here, $\beta$ acts as minus one on $S^1\times\R^+$, fixing two points, while $\alpha$ acts with a minus sign on $S^1_{\varepsilon_i}\times S^1_{\zeta_i}$. 

On the other hand, $V_-$ is curved and given by
\begin{equation}\label{eq:Vminus}
V_- =  (\left( S \setminus E \right) \times S^1_{x_2} \times S^1_{x_3})/\Z_2^\alpha\, ,
\end{equation}
where $S$ is, in fact, the rational elliptic surface $X/\Z_2^\beta$.

To decompose
\begin{equation}
M = \left( X \times S^1_{x_1} \times S^1_{x_2} \times  S^1_{x_3} \right)/ \Z_2^\alpha \times \Z_2^\beta
\end{equation}
as a twisted connected sum, we proceed in a step-wise fashion. Let us first consider the threefold (remember that $\beta$ leaves $S^1_{x_3}$ untouched)
\begin{equation}
Z_\beta =  \left( X \times S^1_{x_1} \times S^1_{x_2} \right)/ \Z_2^\beta \, , 
\end{equation}
which we can describe algebraically by the complete intersection 
\begin{equation}\label{eq:Zinalgway}
\left\{\,\, \begin{aligned}
 y^2 & =  x^3 + x w^4 f_4(z) + w^6 g_6(z) \\
 \xi^2 & = z_1 z_2 \prod_{i=1}^4 (u_1 - \alpha_i u_2)
\end{aligned} \right. \:.
\end{equation}
The fixed point set of $\beta$ is given by the $\mathbb{C}^2/\Z_2$ singularities of the quotient \eqref{eq:Zinalgway}, that are at the eight 2-tori $\xi=z_k=u_1-\alpha_iu_2=0$.

The coordinates $[z_1:z_2]$ span a $\P^1$ on $Z_\beta$.\footnote{
Although $z_1$ and $z_2$ are allowed to vanish simultaneously in the ambient space, where they form the weighted projective space $\P^2_{112}$ together with the coordinate $\xi$, $z_1=z_2=0$ forces also $\xi=0$ by \eqref{eq:Zinalgway}, so that this locus does not exist on $Z_\beta$.} On this $\P^1$ there are $14$ special points: the two points $z_1 z_2=0$, and the $12$ roots of $\Delta(z) = 27g^2 + 4f^3$. We now decompose the $\P^1$ $[z_1:z_2]$ into two patches $P_+$ and $P_-$, such that $z_1 z_2 =0$ are in $P_+$ and the 12 points $\Delta = 0$ are in $P_-$. $P_+$ and $P_-$ are isomorphic to $\C$ and overlap along a circle times an interval.
This gives rise to a decomposition of $Z_\beta$ into two halves $Z_+$ and $Z_-$. 

It is crucial that $\alpha$ acts on both halves separately, without mixing them. 
So, we obtain a decomposition of the orbifold $M$ into two halves 
$$M_+=(Z_+ \times S^1_{x_3})/\Z_2^\alpha \qquad \mbox{ and } \qquad M_-=(Z_- \times S^1_{x_3})/\Z_2^\alpha.$$
They realise $M$ as a twisted connected sum, as we now prove.
We need to show that each half $M_\pm$ can be written as the product of a circle times $V_\pm$. The identification of the circle~$S^1_{w\pm}$ will allow to find the threefold $V_\pm$.

Let us first consider the non-compact three-fold $Z_+$. By assumption, the elliptic curve $E$ described by the first equation in \eqref{eq:Zinalgway} is such that $\Delta$ has no roots for any of the values of $[z_1:z_2]$ in $Z_+$, so that it becomes effectively constant. Moreover, $\beta$ does not act on $E$, so we can write
\begin{equation}
 Z_+ = E \times \left(  P_+ \times S^1_{x_1} \times S^1_{x_2} \right) / \Z_2^\beta  \, .
\end{equation}

We can introduce real local coordinates $\varepsilon_r,\, \varepsilon_i$ on $E$ and $\zeta_r, \,  \zeta_i$ on $P_+$. In terms of these coordinates the action of $\Z_2^\alpha \times \Z_2^\beta$ is
\begin{equation}
 \begin{array}{c|ccccccc}
             & \varepsilon_r & \varepsilon_i & \zeta_r & \zeta_i & x_1 & x_2 & x_3 \\
             \hline
  \beta        & + & +   & -   & -   & -   & -   & +   \\
  \alpha       & + & -   & +   & -   & +   & -   & - \\
  \alpha\beta  & + & -   & -   & +   & -   & +   & - 
 \end{array}
\end{equation}
Note that the action on $\varepsilon_r$ is trivial for the whole orbifold group and we can write
\begin{equation}
M_+ = S^1_{\varepsilon_r} \times \left( S^1_{\varepsilon_i} \times P_+ \times  S^1_{x_1} \times S^1_{x_2} \times  S^1_{x_3}\right) / \Z_2^\alpha \times \Z_2^\beta \:.
\end{equation}
Hence we can identify $S^1_{w+}$ with $S^1_{\varepsilon_r}$ and consequently we derive $V_+$ as in \eqref{eq:Vplus}. Notice that in order for $\alpha$ and $\beta$ to be compatible with the complex structure we need to choose the complex coordinates $x_3+i\varepsilon_i$, $x_1+i\zeta_r$ and $x_2+i\zeta_i$  (up to a phase).

In the asymptotic region, that is away from the fixed point locus of $\beta$, the building block $M_+$ hence looks like
\begin{equation}
M_+^{\rm \infty} = S^1_{\varepsilon_r} \times \left( S^1_{x_1} \times \mathbb{R}_{\zeta_r}^+ \times  \left( S^1_{\varepsilon_i} \times S^1_{\zeta_i} \times S^1_{x_2} \times  S^1_{x_3}\right) / \Z_2^\alpha \right)\:,
\end{equation}
i.e. it has the form we want in the asymptotic region, where the $K3$ surface $\mathcal{S}_+$ has the $T^4/\Z_2$ orbifold form, with complex structure compatible with choosing the complex coordinates $x_3+i\varepsilon_i$ and $x_2+i\zeta_i$  (up to a phase).

Let us now discuss $Z_-$ and $M_-$. On $P_-$, we are away from the two fixed points of $\beta$ at $z_1 z_2 = 0$. Hence, $Z_-$ can be written as the product
\begin{equation}
Z_- =  \left( S \setminus E \right) \times S^1_{x_1} \times S^1_{x_2}\, ,
\end{equation}
where $S$ is the rational elliptic surface described by the first equation in \eqref{eq:Zinalgway}.
Since $\alpha$ acts trivially on $S^1_{x_1}$, the manifold $M_-$ can be described as
\begin{equation}
M_- = S^1_{x_1} \times  \left( \left(S \setminus E \right) \times S^1_{x_2} \times S^1_{x_3} \right)/ \Z_2^\alpha \, .
\end{equation}
We can then identify $S^1_{w-}$ with $S^1_{x_1}$ and $V_-$ with \eqref{eq:Vminus}. The complex structure of $V_-$ tell us that $x_1+ix_2$ must be a complex coordinate.

In the asymptotic region, away from the points where the elliptic curve $E$ degenerates, the rational elliptic surface  becomes the product of a cylinder with an elliptic curve $E=S^1_{\varepsilon_r}\times S^1_{\varepsilon_i}$. In this region the coordinates $\varepsilon_{r,i}$ and $\zeta_{r,i}$ used above are also good coordinates to describe $M_-$, that is
\begin{equation}
M_-^{\rm \infty} = S^1_{x_1} \times \left( S^1_{\varepsilon_r} \times \mathbb{R}_{\zeta_r}^+ \times  \left( S^1_{\varepsilon_i} \times S^1_{\zeta_i} \times S^1_{x_2} \times  S^1_{x_3}\right) / \Z_2^\alpha \right)\:,
\end{equation}
that has the appropriate form. The complex coordinates on  the $K3$ surface $\mathcal{S}_-=T^4/\Z_2$ are now $x_1+ix_2$ and $\varepsilon_i+i\zeta_i$  (up to a phase).

The matching is now complete, as manifestly $S^1_{w\pm}=S^1_{v\mp}$ and the $K3=T^4/\Z_2$ is the same on both sides; to match the latter one needs to do an hyper-K\"ahler rotation (i.e. a Donaldson matching) since the complex structures on the K3 are different in $V_+$ and $V_-$. As we already know that $M$ is a $G_2$ manifold and that $S^1_{w\pm}=S^1_{v\mp}$, this hyper-K\"ahler rotation must be the one used in the TCS construction.

\subsection{Associative Submanifolds}

Having constructed the $G_2$ manifold $M$ as a TCS, we can now describe the associative submanifolds found in Section \ref{sect:g2associatives} in this picture. The upshot of the above section is that we can use 
$\zeta_r$, $\zeta_i$, $\varepsilon_r$, $\varepsilon_i$, 
$x_1$, $x_2$ and $x_3$ as coordinates on (a covering of) the region $M_+ \cap M_-$, and that $x_1$, $\zeta_r$ and $\varepsilon_r$ become coordinates on the base of the K3 fibration of the TCS and $\varepsilon_i,\zeta_i,x_2$ and $x_3$ become coordinates on the K3 fibre.

Let us first consider the associatives $\tilde{C}_\gamma$, which are the product of $S^1_{x_3}$ and sections of an elliptic fibration on $X$ in a complex structure in which $\beta$ acts holomorphically. It follows that we can use $\zeta_r$, $\zeta_i$ and $x_3$ as coordinates on $\tilde{C}_\gamma$ in the region $M_+ \cap M_-$. This means that we can think of these cycles as a fibration of a real two-dimensional manifold (which must be a two-sphere) over an interval in the $\zeta_r$ direction. At the ends of the interval, the fibre collapses, so that we find a compact three-manifold with the topology of a three-sphere. This is precisely the description found in \cite{Braun:2018fdp} for a closely related geometry, where a cycle of this form was constructed for every lattice element in $E_8 \oplus E_8$ and conjectured to be an associative. In particular, it was found in \cite{Braun:2018fdp} that the two $E_8$'s were associated with the two building blocks $V_\pm$: on each end, an isomorphism between $E_8$ and the different ways in which the interval can terminate was found. In the present setup, $V_+$ is a flat orbifold, which leads to a unique choice on $V_+$, while retaining an $E_8$ worth of options on  $V_-$. It was furthermore conjectured in \cite{Braun:2018fdp} that these associatives are sections of a co-associative fibration of $M$ by four-tori $T^4$.\footnote{Such fibrations play the role of a SYZ fibre for a class of mirror maps for type II theories on $G_2$ manifolds~\cite{Braun:2017csz}.} As the $\tilde{C}_\gamma$ become sections of an elliptic fibration on $S\setminus E$ in $V_-$ (in an appropriate complex structure), the fibre of which descends from $\tilde{F} = e^3$ on $X$, we can recover this description with the $T^4$ fibre of $M$ being given by $\tilde{F} \times S^1_{x_1} \times S^1_{x_2}$.

Let us now investigate the cycles $C_\gamma$, which are the product of $S^1_{x_1}$ and sections of an elliptic fibration on $X$ in a complex structure in which $\alpha$ acts holomorphically. In the region $M_+ \cap M_-$, we can hence use $\zeta_r, \varepsilon_r$ and $x_1$ as coordinates on $C_\gamma$. This means that the $C_\gamma$ are in the direction of the $S^3$ base of the K3 fibration with fibres $\mathcal{S}_\pm$ on $M$ and we should think of them as associative sections of this fibration. These associatives are hence fundamentally different from the ones found in \cite{Braun:2018fdp}.

In Section \ref{sect:mainexample}, we found that our orbifold description of $M$ is symmetric under the exchange of $U_1 \leftrightarrow U_3$ and $\alpha \leftrightarrow \beta$. In the light of the discussion of this section, this means that there is in fact a second TCS decomposition of $M$ such that now the $C_\gamma$ correspond to the associatives found in \cite{Braun:2018fdp} and the $\tilde{C}_\gamma$ become sections of a K3 fibration. Ultimately, this stems from the fact that the K3 surface $X$ has multiple calibrated fibrations by $T^2$. It would be very interesting to investigate this further and in particular determine the fate of the various calibrated fibrations once we deform $M$ away from the orbifold limit we are working in. The straight-forward lift of calibrated cycles in $X$ to associatives of $M$ is made possible by the fact that, up to a twist introduced by the orbifolding, $X$ is constant over $T^3/\Z_2\times \Z_2$. As this is expected to be no longer the case once we deform $M$, we do not expect both the $C_\gamma$ and the $\tilde{C}_\gamma$ to persist in a smoothed version of $M$. 

\section{Discussion and Future Directions}

{\it The superpotential}: In section \ref{sect:g2associatives} we constructed an associative 3-sphere in the $G_2$-orbifold $M$ for each element of the two $E_8$-lattices. We expect that each such 3-sphere contributes to the superpotential of the effective four dimensional supergravity theory \cite{Harvey:1999as}. 
This superpotential is naively given by an $E_8$-theta function, following \cite{Donagi:1996yf}. Let us briefly recall how that arises here. The volume of each of the $C_\gamma$ is given by 
\begin{equation}
Vol(C_\gamma) = \int_{C_\gamma} \varphi \:.
\end{equation}
In $M$-theory, due to supersymmetry, there is a natural complexification of the moduli space and the local complex coordinates on the moduli space are given by periods of the complex 3-form, $\Phi = C + i \varphi$, where the $M$-theory $C$-field takes values in the torus $H^3(M, \R)/H^3(M, \Z)$, hence we consider the complex variables,
\begin{equation}
\tau_\gamma = \int_{C_\gamma} \Phi = t_\gamma + i Vol(C_\gamma)\, .
\end{equation}
The contribution of $C_\gamma$ to the superpotential is $W \sim e^{2\pi i \tau_\gamma}$ and therefore, if we assume that each $M$2-brane instanton contributes with the same coefficient, we have
\begin{equation}
W = \sum_{\gamma \in E_8} e^{2\pi i\tau_\gamma}\,.
\end{equation}
Since the $E_8$ lattice has rank eight, this is really a holomorphic function of eight independent complex variables. Denoting by $\tau_i$ the eight variables corresponding to the simple roots and regarding $W$ to be a function of the $\tau_i$, $W$ is in fact the $E_8$-theta function. Similarly, one obtains a second $E_8$ theta function, from the second set of associatives constructed in section two. Recently, there has been significant interest in the general problem(s) of {\it counting associatives} and the related problem of $G_2$-instantons in the mathematics literature \cite{donaldson2009gauge, Joyce:2016fij, haydys2017g2, doan2017counting}. Perhaps the model discussed here and in more general examples could be useful for addressing some of the issues discussed in those papers.

{\it Remark on the moduli dependence of $W$}: Since every moduli chiral multiplet contains an axion field, the entire superpotential of $M$-theory on a $G_2$-holonomy space is non-perturbative. This is {\it not} the case in $F$-theory, where, in principle, the $M$5-brane contributions to $W$ can be multiplied by functions of the complex structure moduli of D3-brane moduli. One should bear this in mind in attempting to identify a dual $F$-theory model, since it is known that non-trivial prefactors depending on complex structure moduli and brane moduli can occur (see e.g. \cite{Ganor:1996pe}).

{\it The scalar potential}: The superpotential plays an important role in determining the vacuum configurations of the effective low energy theory. Combined with the K\"ahler potential for $M$, 
\begin{equation}
K = -3 ln \int_M {1 \over 7} \varphi \wedge *\varphi \:,
\end{equation}
local supersymmetry implies the existence of the following potential function on the moduli space:
\begin{equation}
V = e^{K} (||DW||^2 -3 |W|^2)
\end{equation}
where $DW = \partial W + \partial K W$ and $||DW||^2$ is the natural norm using the inverse of the K\"ahler metric associated with $K$.
It would be very interesting to investigate the properties of the function $V$, given the formulae for $W$ and $K$. For instance, one expects $V$ to have isolated critical points and it would be interesting to investigate e.g. their number theoretic properties in the moduli space.

{\it Different TCS decompostions}: Another interesting observation is that our model orbifold $M$ has two different decompositions as a twisted connected sum. Whilst this may be a consequence of our example being an orbifold, it is apparently the only known example where a TCS decomposition is not unique. This should have interesting consequences for the $G_2$ mirror maps which were discovered for TCS $G_2$ manifolds in \cite{Braun:2017ryx,Braun:2017csz}. Furthermore, the existence of {\it different} TCS decompositions implies the existence of {\it different} K3 fibrations, implying dualities between {\it different} compactifications of heterotic string theory. 

Finally it would be interesting to investigate more general examples constructed with different involutions \cite{2017arXiv170709325J}.

\section*{Acknowledgements}

We would like to thank Lorenzo Foscolo, Mark Haskins, Dominic Joyce and Jason Lotay for discussions. APB thanks the ICTP Trieste for hospitality during various stages of this project. The work of APB is supported by the ERC Consolidator Grant 682608 ``Higgs bundles: Supersymmetric Gauge Theories and Geometry (HIGGSBNDL)''. The work of BSA and ESS is supported by a grant from the Simons Foundation ($\#$488569, Bobby Acharya).

\appendix 

\section{Explicit Representations through String Junctions}
\label{app:StringJunctions}
In this appendix we briefly describe the two-cycles $\Sigma_\gamma$ of the type IIA $K3$ surface $X$ in terms of string junctions. In the complex structure we chose in section~\ref{sec:TypeIIA}, the involution $\beta$ acts holomorphically. We will see explicitly how the two-spheres $\Sigma_\gamma$ are inverted under $\beta$ as sub-manifolds. We will use a detailed description of Hanany-Witten rules \cite{Hanany:1996ie}; the interested reader is referred to e.g. \cite{DeWolfe:1998zf, DeWolfe:1998pr, Hayashi:2010zp}, with references therein for more details. 

The $K3$ surface $X$ is elliptically fibred. 
In the language of section~\ref{sec:Associatives}, the elliptic fibre is in the class $\tilde F= e^3$, while the base is in the class $\tilde \sigma_0=e_3-e^3$. In this picture, the involution $\beta$ can be visualised as a shift of 180 degrees along the azimuth angle in the $S^2$ base of the elliptic fibration. The involution $\beta$ fixes the north pole and the south pole on such an $S^2$. Correspondingly, the fixed point locus in $X$ is give by the two fibres over the two poles, i.e. it is given by two 2-tori as required by the $(10,8,0)$ involution. 

The elliptic fibration degenerates at certain loci of the base. We consider string junctions in the $S^2$ base, starting and ending at these loci (possibly multi-pronged) and we fibre over them the collapsing circle in the 2-torus fibre. This gives rise to irreducible cycles in $X$. Together with $F$ and $\tilde{\sigma}_0$, they generate the full lattice $H^2(X,\Z)$. The string junctions can be added and subtracted following Hanany-Witten rules to make new irreducible cycles such as the $\sigma_\gamma$'s. 

The points of the $S^2$ base where the fiber degenerates are mapped to each other by the involution $\beta$ (that must be a symmetry of the $K3$ surface $X$).\footnote{Unless for the special case in which they coincide with either the north or the south pole.} We can use this fact to sub-divide the base into an eastern and a western hemisphere. The degeneration pints will be half in the first and half in the second hemisphere.
Note  that if we quotient $X$ by $\beta$ we get a rational elliptic surface, and each hemisphere is projected to the $S^2$ base of such elliptic fibration. 
It is hence clear that both the western and the eastern hemisphere contain degeneration points that are enough to generate the full 
$E_8$ lattice of two-cycles by the string junction method. The string junctions making up the classes of the $U_1$ and $U_2$ sublattices must be odd under $\beta$. 

\begin{figure}[H]
\begin{center}
  \includegraphics[height=5cm]{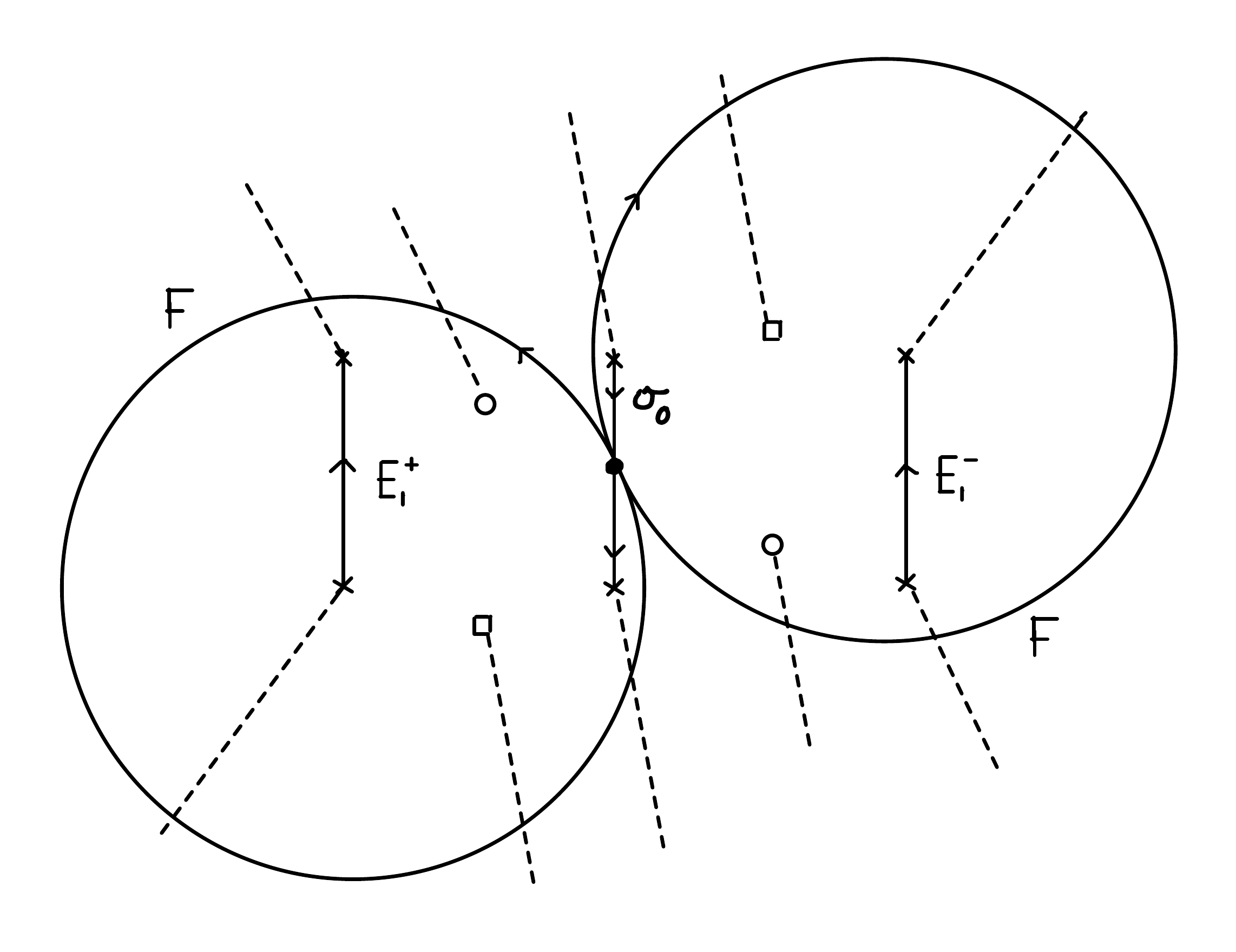}
 \caption{\label{fig:SumJunctionsStart} The sum of cycles $\sigma_\gamma$ given by equation \eqref{eq:junctionEx} represented as a sum of string junctions. The circle and square represent additional unspecified degeneration loci on the $K3$ base. The black dot represents the orbifold fixed point.}
\end{center}
\end{figure}

In order to illustrate how this construction works, we now describe a string junction corresponding to an odd irreducible 2-cycle under $\beta$. Such a junction must cross at least one of the poles at least once. Further, its behaviour on the western hemisphere must be reflected in the behaviour on the eastern hemisphere. That is, if a string junction start from a point in the west, it must end at the image point in the east, and the fibre $S^1$ degenerates in the same manner at the two points. Take for example the cycle 
\begin{equation}
\label{eq:junctionEx}
\sigma=\tilde \sigma_0+2F+E_1^+ + E_1^-\:,
\end{equation}
where $E_1^+$ ($E_1^-$) is a simple root of $E_8$ ($E_8'$), represented by a two-sphere, i.e. $(E_1^{\pm})^2=-2$. 
In Figure \ref{fig:SumJunctionsStart} we draw the building blocks: The class $F$ is represented by a great circle which crosses both the north and south pole (it wraps one of the circles in the 2-torus fibre). Note that $F$ is naturally odd under  $\beta$. Being an odd class, the junction representing $\sigma_0$ must cross one of the poles and must wrap the other circle in the fibre in order to intersect with $F$ once. We have also drawn the classes $E_1^\pm$, which are represented by strings emanating and ending on degeneration loci of the same type. Note that $E_1^\pm\rightarrow -E_1^\mp$ under $\beta$.

\begin{figure}[H]
\begin{center}
  \includegraphics[height=5cm]{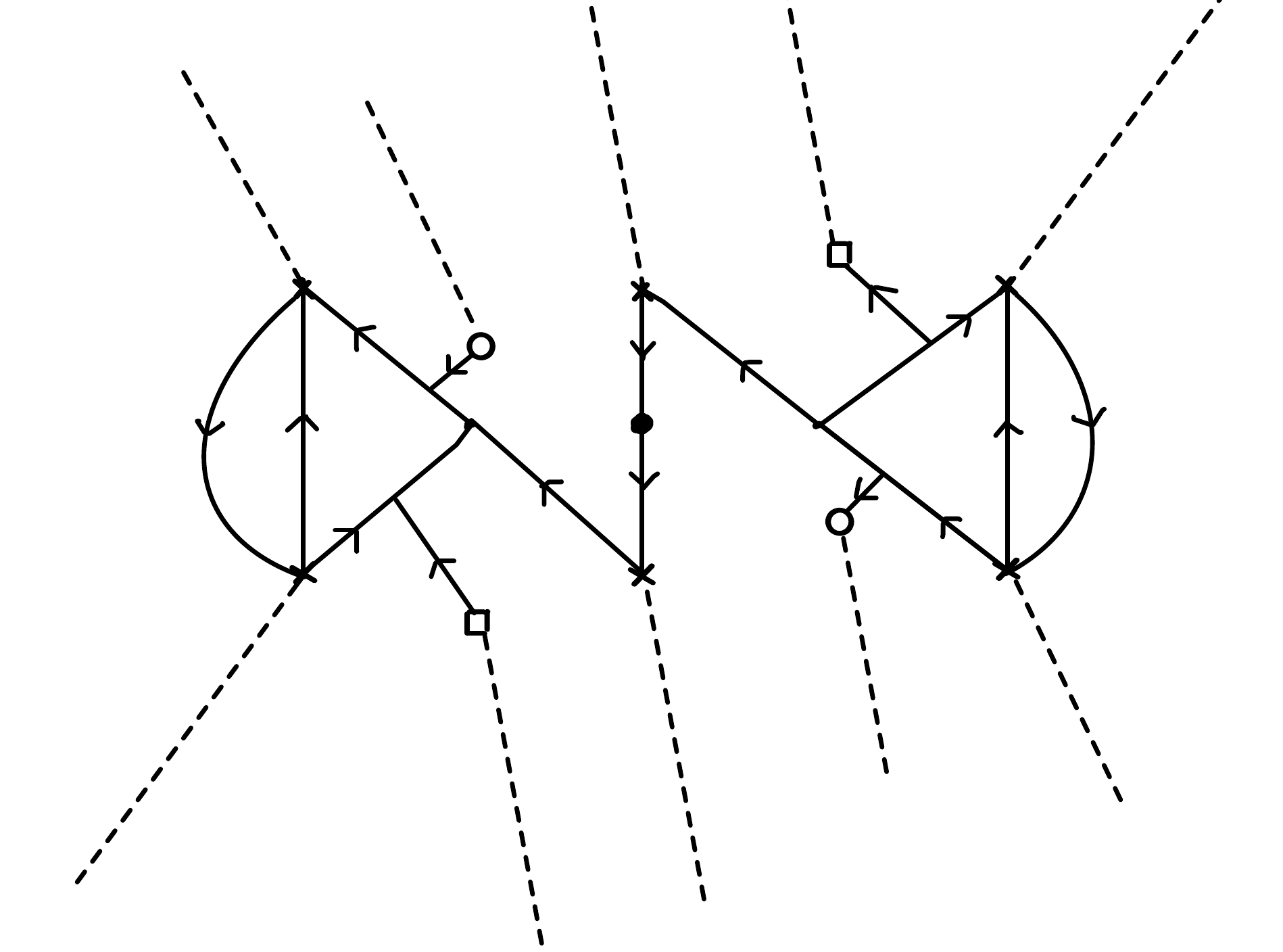}
 \caption{\label{fig:SumJunctionsMiddle} An intermediate step towards the irreducible representative of the string junction for $\sigma$.}
\end{center}
\end{figure}

\begin{figure}[H]
\begin{center}
  \includegraphics[height=5cm]{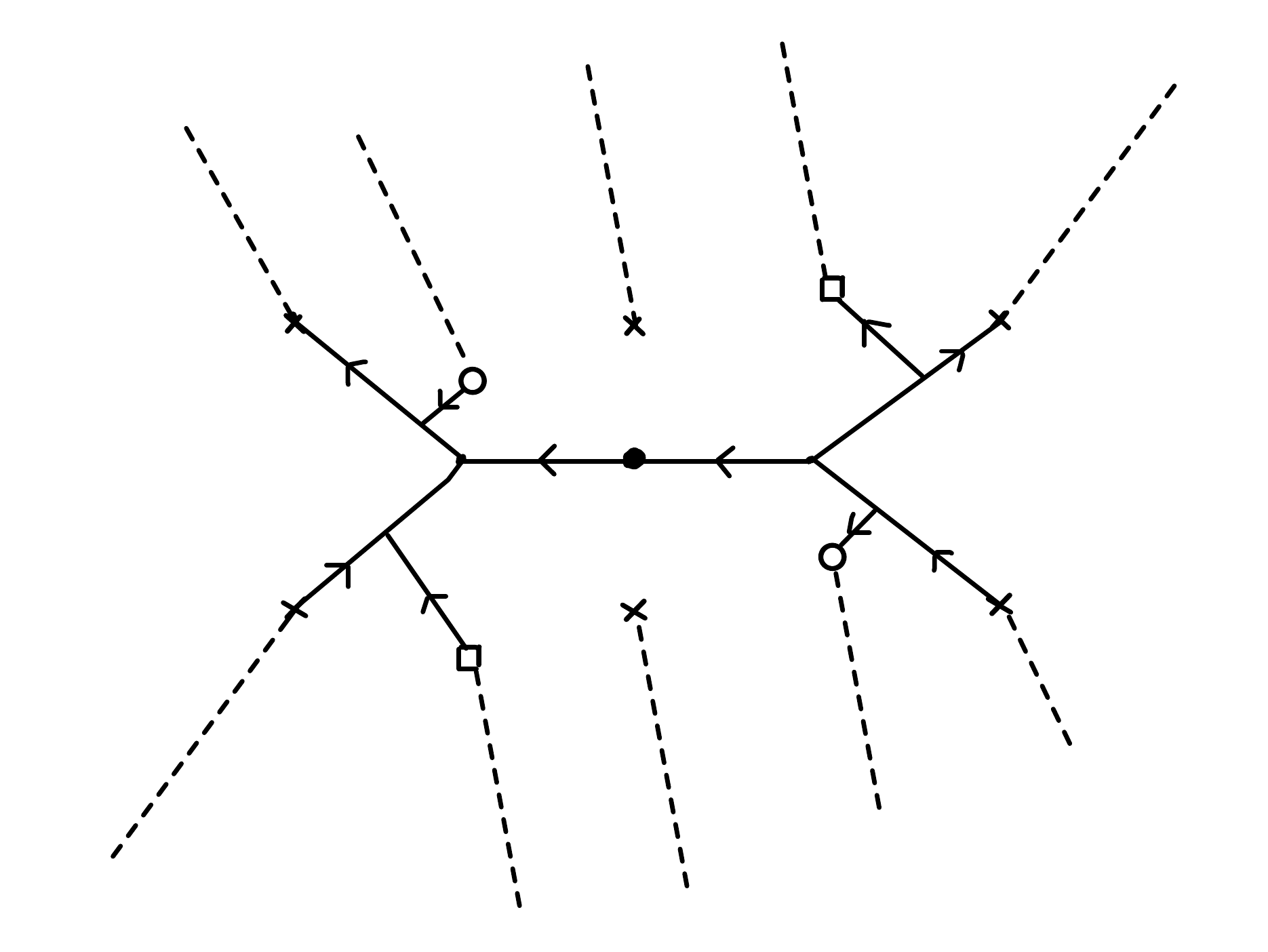}
 \caption{\label{fig:SumJunctionsFinal} The class $\sigma$ represented as a single string junction corresponding to a two-sphere embedded in the $K3$ surface.}
\end{center}
\end{figure}

The class $\sigma$ cointains the 2-cycle built on the union of the string junctions in Figure~\ref{fig:SumJunctionsStart}. This cycle is manifestly odd. We now deform this cycle to obtain a connected representative.
Following Hanany-Witten, one can perform the same steps on both western and eastern hemispheres, in order to keep the total string junction odd at all times. An intermediate step is drawn in Figure \ref{fig:SumJunctionsMiddle}, and the final string junction is shown in Figure \ref{fig:SumJunctionsFinal}. As claimed, this junction is odd under $\beta$.

\providecommand{\href}[2]{#2}\begingroup\raggedright\endgroup

\end{document}